# Towards indicating interdisciplinarity: Characterizing interdisciplinary knowledge flow

Hongyu Zhou[1], Raf Guns[1], and Tim C. E. Engels[1]

[1] *Hongyu.Zhou@uantwerpen.be, Raf.Guns@uantwerpen.be, Tim.Engels@uantwerpen.be*
Centre for R&D Monitoring (ECOOM), Faculty of Social Sciences, University of Antwerp
Middelheimlaan 1, 2020 Antwerp (Belgium)

# Towards indicating interdisciplinarity: Characterizing interdisciplinary knowledge flow

## ABSTRACT

This study contributes to the recent discussions on indicating interdisciplinarity, i.e., going beyond catch-all metrics of interdisciplinarity. We propose a contextual framework to improve the granularity and usability of the existing methodology for interdisciplinary knowledge flow (IKF) in which scientific disciplines import and export knowledge from/to other disciplines. To characterize the knowledge exchange between disciplines, we recognize three aspects of IKF under this framework, namely, broadness, intensity, and homogeneity. We show how to utilize them to uncover different forms of interdisciplinarity, especially between disciplines with the largest volume of IKF. We apply this framework in two use cases, one at the level of disciplines and one at the level of journals, to show how it can offer a more holistic and detailed viewpoint on the interdisciplinarity of scientific entities than aggregated and context-unaware indicators. We further compare our proposed framework, an indicating process, with established indicators and discuss how such information tools on interdisciplinarity can assist science policy practices such as performance-based research funding systems and panel-based peer review processes.

## 1. INTRODUCTION

How scientific disciplines import and export knowledge from/to others defines their role in the scientific community and informs their values and actions on interdisciplinarity. Understanding interdisciplinarity from the perspective of knowledge flow (linkage) between disciplines is steeped in the history of quantitative science studies. In 1978, Garfield, Malin, and Small (1978) pointed out that interdisciplinary activity can be studied using "linkages between specialties of diverse subject matter." Their main focus back then was to discover specialties of science and linkages among them to arrive at a "map of science". The study of interdisciplinarity of a certain discipline (or other entities, e.g. researchers or institutions) has bloomed since the 1980s, after which many empirical indicators of interdisciplinarity have been proposed and employed.

The trajectory of quantifying interdisciplinarity can be divided into two phases. Indicators proposed in the first phase (roughly before 2000) were based on cross-disciplinary citations (knowledge flow) to infer the exchange of knowledge/resources. According to Rafols & Meyer (2010), the percentage of citations outside categories (PCOC), proposed by Porter & Chubin (1985), was the "most common indicator of interdisciplinarity". The "category" here is normally operationalized as a set of journals to represent a discipline. For instance, van

Leeuwen & Tijssen (2000) investigate cross-disciplinary citation links between 119 disciplines using PCOC and identified disciplines such as meteorology and atmospheric sciences as being more interdisciplinary. Another important indicator in this phase is the percentage of citations towards category (PCTC), which has a longer applied history and has attracted a broader audience in various fields. As early as 1952, Broadus (1952) analyzed the percentage of articles from different subjects (disciplines) cited in the American Sociological Review in 1950. Since then, this indicator is employed by researchers from various disciplines and topics, such as linguistics (Rappaport, 1971), marketing (Goldman, 1979), consumer research (Leong, 1989), agricultural education (Radhakrishna, 1992), and social science in general (Rigney & Barnes, 1980). It is typically referred to as citation analysis or citation study by general researchers, whose primary goal is to unveil the knowledge constitution of a certain discipline or journal, instead of quantifying the level of interdisciplinarity directly. Although simple and intuitive, the two indicators in the first phase are versatile in delivering contextual understandings of interdisciplinarity and are still utilized (directly or in modified form) in recent studies (Angrist et al., 2020; Huang et al., 2022; Truc et al., 2020).

The second phase of quantifying interdisciplinarity is marked by the introduction of the diversity framework (Stirling, 2007). Since 2000, some researchers have started to operationalize interdisciplinarity as cognitive diversity, or components of diversity, for instance, diversity of relationships with other disciplines using the Pratt index (Morillo et al., 2003). A milestone article in this phase is published by Rafols & Meyer (2010), in which a framework to quantify interdisciplinarity using diversity and its three components are explicitly proposed. From then on, a significant proportion of research on interdisciplinarity adopt this framework and different variants of indicators are proposed (Leydesdorff et al., 2019; Rousseau, 2018; Zhang et al., 2016). The main objective of research in this phase is to find an ideal aggregated indicator to gauge interdisciplinarity for various purposes.

However, recently some researchers voiced concerns about the current interdisciplinarity measures through both empirical analysis and theoretical discussions. Wang & Schneider (2020) survey 16 existing interdisciplinarity measures in an empirical study and concluded that the current measurements are "both confusing and unsatisfying" due to a lack of consistency in results. They encourage "something different" instead of "more of the same" in methodology to measure "the multidimensional and complex construct of interdisciplinarity" (Wang & Schneider, 2020, p. 239). In a similar spirit, Rafols (2020b) discusses the failed efforts of universal indicators and suggests the notion of "indicating interdisciplinarity", which is "a contextualized process … indicating where and how interdisciplinarity develops as a process". Marres and de Rijcke (2020) describe "indicating" as "participatory, abductive, interactive, and informed by design" and propose digital mapping as a promising tool to indicate interdisciplinarity (p. 1042).

At the center of these debates is the complexity of understanding interdisciplinarity and the difficulty in operationalizing the concept. We agree with the recent concerns that the multidimensionality of interdisciplinarity cannot be entirely represented by a binary label (interdisciplinary or disciplinary) or a single value in a continuous range (more or less interdisciplinary). The interdisciplinarity of certain entities may be more accurately and comprehensively described through their interdisciplinary engagements with all disciplines (vectors), without aggregations to unidimensional indicators (scalars). In this article, we would like to further develop an existing indicator on characterizing interdisciplinary knowledge flow (i.e., PCTC from the first phase of indicators on interdisciplinarity) by offering more nuances of interdisciplinary knowledge flow and contributing to the literature a method that adopts the idea of "indicating interdisciplinarity".

## 2. THE UNEXPLORED ASPECTS OF INTERDISCIPLINARY KNOWLEDGE FLOW

As mentioned in the previous section, a popular indicator of characterizing interdisciplinary knowledge flow is the percentage of citations towards category (PCTC). For a certain entity of analysis, one can calculate the percentage of citations from it to other disciplines which demonstrates its relative preference for interdisciplinary engagements. The results are normally presented with tables or with line plots showing temporal trends. PCTC is a useful indicator in presenting the knowledge portfolio of research entities and delivers clear information on interdisciplinary knowledge borrowing.

Nonetheless, some important aspects of interdisciplinary knowledge flow are not well captured using PCTC, which leaves room for further improvement. We discuss two scenarios that PCTC fails to differentiate.

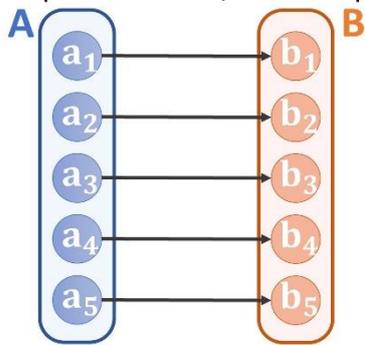
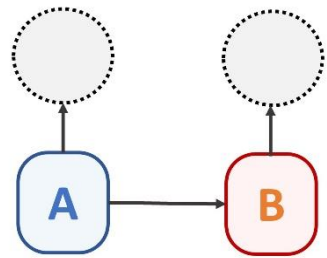
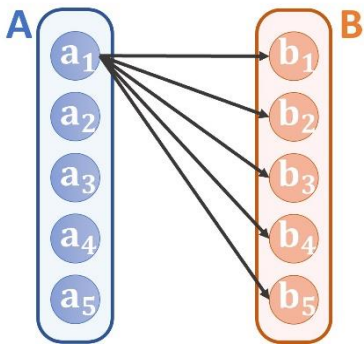
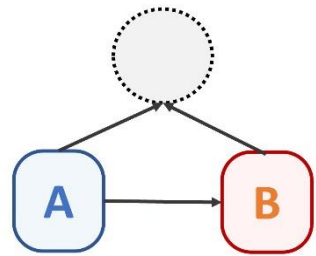

**Figure 1. Aspects of interdisciplinary knowledge flow not captured by PCTC.**

Scenario 1 (Figure 1 a&b): Consider two disciplines A and B, each consisting of five publications with ten references, and five citations are made from articles in A to B. With an equal amount and percentage of citations from A to B, all five publications of A cite B in Figure 1a, whereas only one A publication, $a_1$, cites B in Figure 1b. In the former case (Figure 1a), we consider B exerts a **broad** impact on A, since all publications of A cite B. In the latter case (Figure 1b), we consider B exerts a narrower impact on A, since only one A publication out of five is influenced by B. However, the influence is quite **intense** since half of this publication's references come from B. We argue that in this case, B exerts a narrow but intense impact on A. When analyzing interdisciplinary relationships, in addition to the volume of interdisciplinary citations, such details on citation flow are indispensable in characterizing the nature of knowledge flow between fields, yet they are lacking in PCTC.

Scenario 2 (Figure 1 c&d): Suppose one observes citations between disciplines A and B, on what basis are interdisciplinary citations between them made? It may be that A cites B since they start from similar knowledge bases, and citations are triggered by knowledge proximity or homogeneity. Or perhaps A cites B since B is a different and remote knowledge cluster, with the goal of borrowing knowledge that it cannot access in its own knowledge base. Both rationales are possible and deserve to be differentiated. For the example in Figure 1c, two disciplines cite each other yet exhibit no overlap in their knowledge bases. It shows that the observed citations are made between two **heterogeneous** entities. In the example of Figure 1d, the two disciplines share the exact same knowledge base and citations between them. It shows that the observed citations are made between two **homogeneous** entities. One may expect that the more two entities are cognitively similar, the more they will cite each other. The alternative may deserve more attention that uncovers the "unexpected" interdisciplinary citations between heterogeneous entities. Nonetheless, we believe when describing the interdisciplinary knowledge flow between entities, the cognitive similarity between them, similar to the notion of disparity in Rafols & Meyer (2010), should also be studied as an important aspect that PCTC fails to cover.

The elements discussed in the two scenarios should be scrutinized when deciphering the interdisciplinary knowledge flow of research entities, which may contribute to a more in-depth and multi-faceted understanding of interdisciplinarity. In this paper, we hope to propose three indicators that can be utilized as a three-step process to capture the unexplored aspects of interdisciplinary knowledge flow.

## 3. OPERATIONALIZATION

To quantify the discussed aspects of interdisciplinary knowledge flow, we propose three indicators, namely, broadness, intensity, and homogeneity. The first two account for the first scenario in Figure 1 and quantify the shape of citations, supplementing the volume of citations. The third indicator describes the cognitive similarity between the citing and cited entities. We use the three indicators to describe the interdisciplinary relationships between two entities of interest. The citing and cited entities can be either the same type of entities such as fields, or they can be different types of entities, e.g., the citing entities are journals while the cited entities are disciplines. We now introduce them one by one.

*3.1 Broadness*

The broadness ($B$) of interdisciplinary knowledge flow quantifies how broad the influence of the cited discipline on the citing entity is. The broadness of entity $X$ citing discipline $D$, $B(X, D)$, is defined as the percentage of publications in the studied entity that cite another discipline. We consider greater broadness between them when a greater percentage of publications in the citing entity cites the discipline of interest. With equal proportions of outward citations towards two disciplines, an entity may exhibit greater broadness with respect to one of the disciplines since more publications cite it.

To provide a mathematical definition, let us suppose a citation matrix $M$ ($n \times n$), with n denoting the total number of all publications in a given period. For any two publications $i$ and $j$, $M_{ij} = 1$ if $i$ cites $j$, and 0 otherwise. An entity $X$ (e.g., field, journal, or set of publication output of institutions or researchers) is represented by the $n$ publications classified under it: $X = \{x_1, \ldots, x_n\}$. We use $|X|$ to denote the number of publications in $X$. The broadness ($B$) of IKF of $X$ citing $Y$ captures the percentage of papers in entity $X$ that cite entity $Y$. It is defined as:

$$B(X, Y) = \frac{\sum_{i \in X} \delta_i}{|X|} \quad (1)$$

where $\delta_i = \begin{cases} 1, if\ \sum_{j \in Y} M_{ij} > 0 \\ 0, if\ \sum_{j \in Y} M_{ij} = 0 \end{cases}$.

*3.2 Intensity*

After quantifying the broadness of interdisciplinary knowledge flow, we place our focus on a subset of entity $X$ that cite $Y$ at least once, namely $X'$ ($\delta_i = 1$). We want to study, among the publications of $X$ that are influenced by $Y$, how intense is the influence. The intensity of interdisciplinary knowledge flow is, therefore, defined as the percentage of citations from $X'$ to $Y$, equivalent to citations from $X$ to $Y$, in all the outward citations of $X'$.

The mathematical definition of intensity ($I$) is a modified version of PCTC, with both quantifying the volume of citation links between entities. The calculations of the two follow the same logic with one additional restraint added for intensity, as shown in Equations 2 and 3. PCTC (Equation 3) quantifies the percentage of citations from the citing entity $X$ to the cited entity $Y$ in the entire outward citations made by $X$, regardless of who the recipient entities are in the denominator. The intensity (Equation 2), on the other hand, controls the scope of the denominator by introducing $\delta_i$, which limits the inclusion of outward citations from $X$ only for those who cite Y, i.e., $X'$.

$$I(X, Y) = \frac{\sum_{i \in X, j \in Y} M_{ij}}{\sum_{i \in X, j = 1, \ldots, n}(M_{ij}\delta_i)} \quad (2)$$

$$PCTC(X, Y) = \frac{\sum_{i \in X, j \in Y} M_{ij}}{\sum_{i \in X, j = 1, \ldots, n} M_{ij}} \quad (3)$$

The goal of quantifying broadness and intensity is to describe the shape of citation flow and complement what PCTC fails to inform. For example, suppose we observe a growing proportion of citations between two disciplines (increasing PCTC), is the increase caused by the same amount of citations from more publications, or is it caused by more citations from the same amount of publications, or more citations from more publications? With only PCTC quantifying the volume of citations, one cannot accurately differentiate the three scenarios. Let's consider another concrete example. In Figure 2, the citation relationship between disciplines A and C is shown using a citation matrix. The vertical rows denote citing publications from discipline A ($A_1, A_2, A_3, A_4, A_5$) and C ($C_1, C_2, C_3, C_4, C_5$), whereas the horizontal columns represent the cited publication from discipline B ($B_1, B_2, \ldots, B_{10}$). The last column named "Total" records the total number of outward citations by the corresponding citing publication. If we quantify the two selected relationships, we find that PCTC(A citing B) is equal to PCTC(C citing B), therefore arriving at the same measured volume of knowledge flow for the two pairs. However, under the inspection of our proposed framework, both the broadness and intensity of IKF are assigned different values. The broadness of A citing B is larger than that of C citing B since four out of five papers in total from A cite at least one paper from B, compared to one out of five in the case of C citing B. The intensity of A citing B, on the other hand, is less than that of C citing B since only $C_1$ cites discipline B, but more intensely (with five citations in one paper).

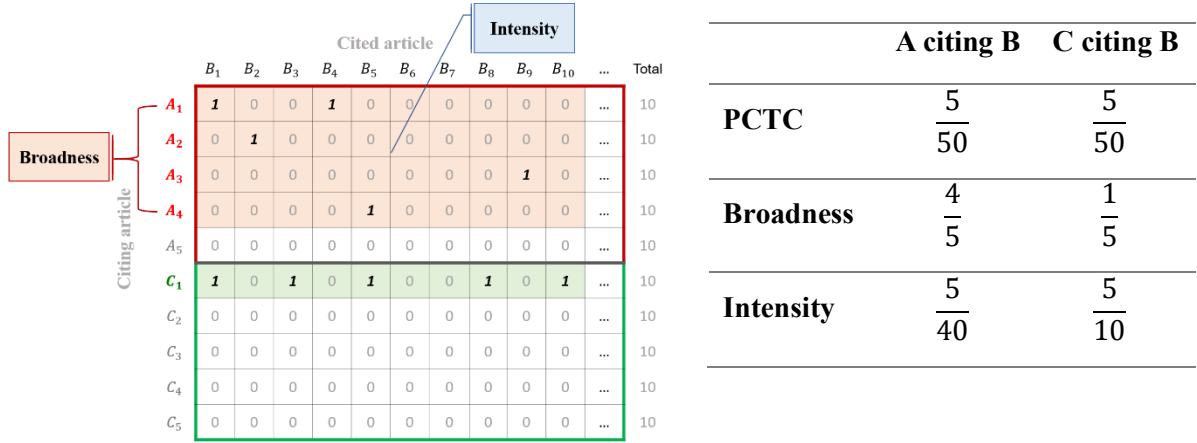

**Figure 2. Demonstration of the framework (broadness and intensity)**

*3.3 Homogeneity*

After observing the direct citations between studied entities and disciplines, we further describe the cognitive similarity between both entities. The homogeneity indicator answers the question: to what extent do the two share a similar knowledge base? We define the homogeneity of X citing Y as the percentage of references of X that are co-cited by Y. If X cites a great proportion of references that Y also utilizes, we say that X exhibits high homogeneity in the knowledge base with Y, and vice versa. It is defined as:

$$H(X,Y) = \frac{\sum_{i \in X, \gamma = 1,\ldots,n} M_{i\gamma} \varphi_{\gamma,Y}}{\sum_{i \in X, j = 1,\ldots,n} M_{ij}} \quad (4)$$

where $\varphi_{Y,Y} = \begin{cases} 1, if \ \sum_{j \in Y} M_{jY} > 0 \\ 0, if \ \sum_{j \in Y} M_{jY} = 0 \end{cases}$.

Note that the denominator of homogeneity indicates the total number of outward citations from *X*, which is exactly the same as PCTC's.

*3.4 Incorporating all three aspects of IKF*

To sum up, we first employ *broadness* to determine the height of a sub-matrix in which only papers having citations from the focal entity to the target discipline are included. For the recognized sub-matrix, we then quantify the *intensity* of citations from the focal entity to the target discipline in the focal entity's complete citation portfolio. Besides empirically representing two distinct axes of the citation matrix between two entities, the two aspects also capture different connotations in discipline-wise relationships. For instance, consider the following two scenarios:

- Scenario 1: $B(X,Y) = 0.9, \ I(X,Y) = 0.1$
- Scenario 2: $B(X,Y) = 0.1, \ I(X,Y) = 0.9$

From the perspective of X, Y plays completely different roles in the two scenarios. In scenario 1, Y contributes to X as a universal ingredient that 90% of publications in X include Y in the knowledge base. But when X cites Y, X only includes 10% of knowledge from Y, which shows that the knowledge borrowing is selective, concentrating on a small number of influential publications from Y. Scenario 1 may correspond to, e.g., the relation between the quantitative branch of social sciences and statistics. The latter may be needed in the majority of quantitative papers, but only a few references are typically sufficient enough to introduce the methodology. In scenario 2, Y only influences a small cluster of X yet rather intensively. This may correspond to how Philosophy influences AI research: only a small local cluster in AI, namely AI ethics, cites Philosophy but rather intensively. To use cooking metaphors to classify the role of Y to X, in scenario 1 Y could be *salt*, which is necessary for most dishes yet in small volumes; in scenario 2, Y could be *turmeric*, which is not pervasive in cooking but used intensively when making curry-based cuisine.

We use the third aspect, homogeneity, to better understand the "nature" of knowledge borrowing, that is, whether the citing and cited entity are more homogeneous or more heterogeneous. We explain the rationale for considering this aspect of IKF with an example of international trade. Suppose that countries X and Y both trade with five countries in the world and have an equal trade-to-GDP ratio. Yet country X trades with five neighboring countries, whereas country Y trades with five countries from five continents. In spite of their equal reliance on international trade, country X can be considered a regional player in the trade that collaborates closely with neighboring countries, whereas country Y operates more like a global player. In this example, in addition to the proportion of international trade, the question with whom a country trades also informs its trade profile. In our study of IKF, we are also curious about entities' cognitive similarity with cited disciplines. One may expect that if two entities are cognitively similar, they are more likely to cite each other. Cases that contradict this expectation may reveal unique patterns of interdisciplinarity, e.g., many citations between cognitively heterogeneous entities, or few citations between cognitively homogenous entities.

*3.5 Towards "indicating interdisciplinarity"*

Rafols (2020b) framed indicating as "indicating where and how interdisciplinarity develops as a process, given the particular understanding relevant to the specific policy goals". We understand the core of this definition is to examine knowledge portfolios and delineate the process of knowledge integration, instead of one-off indicators without context. In addition, Rafols also stressed the importance of directionality in research and innovation and the evaluation of interdisciplinarity. He suggests going beyond values of interdisciplinarity (unidimensional indicators) to indicate "orientations of the research contents" (distributions). We categorize our proposal as indicating interdisciplinarity since it also provides the contexts of diversity in the knowledge base and indicates the exact (disciplinary) directions from which the evaluated entity imports knowledge, by showing their distributions.

For a certain entity of interest, the framework yields three arrays of values $(B_n, I_n, H_n)$ representing each of the three indicators, where the array size $n$ is the number of potential extramural fields from which the entity under scrutiny can import knowledge. Based on the conceptualization of interdisciplinarity discussed above, entities with greater interdisciplinarity are therefore expected to associate with greater broadness (widely influenced by other fields), greater intensity (intensively influenced by other fields), and greater homogeneity in knowledge base with other fields (greater cognitive similarity in the knowledge base with other fields) since they engage more actively in knowledge integration.

A key issue when utilizing the proposed framework for comparisons between entities is whether the indicators are size-dependent. Not all entities – whether they are disciplines, journals, or something else – are equally large, and we want to avoid systematically characterizing larger fields as more interdisciplinary, only because of their size. To put it differently, a random subset of an entity's publications should receive similar values of the IKF indicators. Through mathematical deduction and empirical tests, we show that the three indicators enjoy the properties of monotonicity and size independence and, therefore, prove the validity of making comparisons using the proposed framework. The details are shown in Supplementary Information (SI) B.

We would also like to emphasize that our proposed method cannot and should not provide one-off judgments on interdisciplinarity without clearly defining the context. We suggest a comparative understanding of interdisciplinarity in the sense that one could deduct the relative scale, type, and orientation of interdisciplinarity of an entity with comparisons with its comparable pairs, yet cannot be inferred on its own. Interdisciplinarity is a dynamic process that varies among disciplines and evolves over time. One cannot be classified as interdisciplinary without benchmarks to its disciplinary and temporal norm. Therefore, a second operationalization of this method is proposed as the difference in knowledge integration between a certain entity and a comparative benchmark $(B_n - B'_n, I_n - I'_n, H_n - H'_n)$. If the analyzed entity possesses greater knowledge integration in comparison to the chosen benchmark for a certain indicator, we can say it associates with greater interdisciplinarity in relative terms. An example of such a benchmark comparison is provided in section 5.2.2.

## 4. OUTLINE OF EMPIRICAL STUDY

To test the validity of our proposal, we first examine the distribution of the three aspects of IKF and then study the relationships between the three aspects of IKF using empirical citation data (section 4.1.1). We examine the correlations between every pair of aspects by generating linear fits to their relationships. Discipline pairs that deviate significantly from linear fits are highlighted and discussed through case studies. Besides analysis for all discipline pairs, we look specifically into the IKF between disciplines that possess the 10% highest volume of knowledge flow, i.e. for every discipline, we study the 10% most cited disciplines separately. In a second validation analysis (section 4.1.2), we examine the relationships between the three aspects of IKF and the volume of citations in IKF. Through these two analyses, we aim to demonstrate how the proposed framework can uncover different types of IKF and what kind of discriminative power it can contribute.

We further put the framework into practice with two use cases and illustrate how to indicate and compare the interdisciplinarity of disciplines (direct links with all disciplines) and journals (relative scales in a local discipline setting) using our proposed framework. In the use case of indicating discipline level interdisciplinarity (section 4.2.1), we provide examples for eight disciplines to illustrate how our model can detect their unique patterns of IKF. We also compare results from our framework with existing interdisciplinarity indicators such as True Diversity (Zhang et al., 2016) to show how the proposed multidimensional framework is empirically related to established indicators of interdisciplinarity. In the use case of indicating journal-wise interdisciplinarity in a local discipline setting (section 4.2.2), we select 87 journals from Library & Information Science (LIS) to constitute a sample representing this discipline at large. We then choose seven flagship journals and compare their relative interdisciplinary knowledge preferences with that of LIS in general. We also examine the correlation between a journal's deviation in interdisciplinary knowledge from its affiliated discipline at large and its level of interdisciplinarity globally. The results illustrate how the framework enables a more granular and in-depth understanding of journal-level interdisciplinarity.

We place our focus in this study on the citing side of IKF, which can also be interpreted as knowledge borrowing.

## 5. DATA

The data in this study is harvested from the in-house version of Web of Science (including Science Citation Index Expanded [SCIE], Social Science Citation Index [SSCI], and Arts and Humanities Citation Index [A&HCI]) hosted at ECOOM KU Leuven. We select all citable items (e.g., article, review, and letter) published in 2015 and 2016 from all 74 disciplines recognized in the Leuven-Budapest classification scheme (Glänzel et al., 2016; Glänzel & Schubert, 2003). We investigate 2,995,186 publications and 89,956,048 references made by them. For each discipline pair, we calculate the aforementioned three indicators, as well as the number of citations from the citing discipline to the cited discipline and the interdisciplinarity score of the citing discipline. For the use case of journals, we selected 87 journals in LIS from the Journal Citation Reports 2019 (JCR 2019) and all 8,359 articles published in these journals in 2015 and 2016.

## 6. RESULTS

*6.1 Examining the framework*

6.1.1 The distribution of the three aspects of IKF

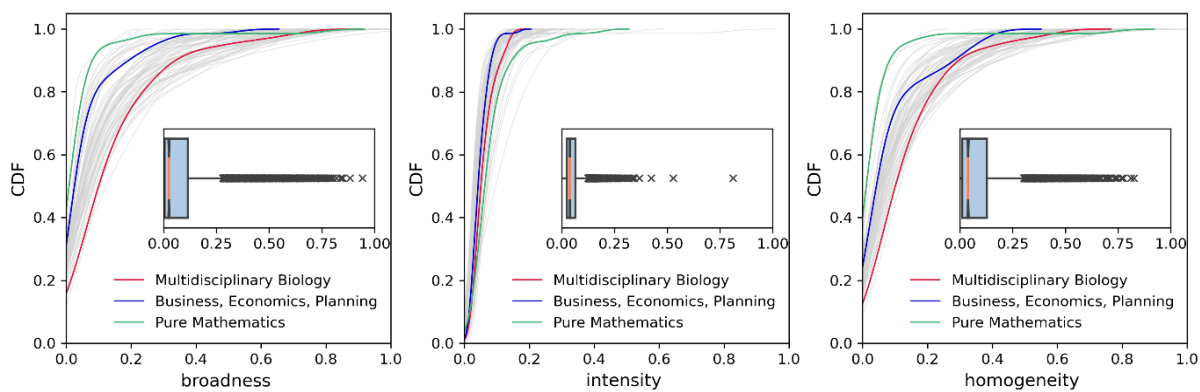

**Figure 3. The distribution of the three aspects of IKF. The main plots present the cumulative distribution function (CDF) of the three aspects for the studied 74 disciplines. Each line denotes the IKF of a discipline, described by broadness, intensity, and homogeneity, respectively. Three disciplines are highlighted with colors. The inset plots present the distribution of the indicators for all discipline pairs.**

We find skewed distributions for all three aspects of IKF (Figure 3), which is indicative that disciplines exhibit greater interdisciplinary engagements with only a few disciplines. Some disciplines exhibit a particularly skewed distribution for broadness and homogeneity, e.g. Pure Mathematics, which indicates their limited interdisciplinary engagement with a small set of disciplines. On the other hand, disciplines such as Multidisciplinary Biology exhibit less skewness since they interact more frequently with more disciplines.

In our analysis of the IKF between disciplines, the value for broadness ranges from zero to a maximum of 0.94 (Cell Biology citing Biochemistry/Biophysics/Molecular Biology). Values for intensity range from zero to a maximum of 0.81 (Literature citing Pure Mathematics, with broadness of 0.0006). It characterizes a relationship where a very small group of publications from Literature utilized references from Pure Mathematics intensively. It also demonstrates the necessity of using intensity together with broadness, instead of only itself, to gain a meaningful understanding of the shape of IKF. For the homogeneity aspect, the value ranges from zero (Organic & Medicinal Chemistry citing Architecture) to 0.80 (Atomic, Molecular & Chemical Physics citing Physical Chemistry).

6.1.2 Relationships among the three aspects of IKF

We further examine the relationships between the three aspects of IKF and present how their combinations may capture different types of interdisciplinary engagements (Figure 4). The $R^2$ values of linear fits are denoted in the plots. It is clear that intensity in particular offers a disparate perspective on IKF from the other two aspects. Its $R^2$ values with both broadness and homogeneity are relatively small (0.214 and 0.253, respectively), which demonstrates once again the uniqueness of intensity. As expected, a clear and relatively strong positive relationship can be found between broadness and homogeneity, i.e. discipline pairs with greater cognitive similarity are associated with broader impact. Nevertheless, they are not isomorphic, either theoretically or empirically. The broadness aspect describes the width of citing behavior, whereas the homogeneity aspect quantifies the degree of overlap in the knowledge base. We can also differentiate them in empirical analyses. In Figure 4b, some special data points are highlighted as case studies since they deviate significantly from the linear fit and the main cluster. The details of these cases are shown in Table 1. For the red dots in Figure 4b, the relationships between the citing and cited discipline possess lower homogeneity and higher broadness than expected. This indicates that although the two disciplines have disparate knowledge bases, knowledge from the cited discipline is still broadly adopted by the citing discipline resulting in a broader diffusion than what one might expect. For instance, in cases 1, 2, and 3 (Figure 4b), Applied Mathematics and Computer Science/Information Technology, two methodology-oriented disciplines, constitute significant knowledge exporters with a broader impact on Genetics & Developmental Biology and Pure & Applied Ecology than average scenarios, despite the substantial cognitive distance between them. Two cited disciplines with multidisciplinary nature in cases 4 and 5 also have broader yet cognitively different impacts on the citing disciplines. Even though their knowledge base is different from that of citing disciplines due to the diversity and span of topics, they also contribute a valuable and broader knowledge source from which citing disciplines can learn. On the other hand, cases 6 and 7 exhibit higher homogeneity and lower broadness than expected. They represent discipline pairs that are built on similar knowledge constructs yet cite each other in a disproportional broadness. In case 6, Architecture shares 61.6% of references with General & Traditional Engineering, however, only 40.4% of its publications cite this discipline. In case 7, History & Archaeology shares 46.5% of the knowledge base with Geosciences & Technology, but the broadness of its citing behavior vis-à-vis Geosciences & Technology is lower than expected. Among other factors, such deviations can partly relate to imbalanced discipline sizes (see SI Figure. A1) and the epistemic characteristics of the disciplines involved (e.g., fundamental versus applied research).

We also observe some general typology of interdisciplinary citations that illustrates the characteristic functions of disciplines. For instance, in Figure 2a, Psychology & Behavioral Sciences (red dots) is cited by two disciplines with higher intensity than expected, which means it contributes intense knowledge to a subgroup of the citing discipline, such as studies relating to Forensic Psychology in Law or Neuroaesthetics in Arts & Design. On the other hand, Applied Physics contributes a broader yet less intense impact on knowledge flow to Physical Chemistry and Polymer Science, which shows that its knowledge serves as a common tool that does not require high maintenance, i.e. complex theoretical underpinning or heavy argument.

We further look into disciplines' pairwise relationships with the highest volume of knowledge flow. For each discipline, based on the number of citations from them to the other disciplines, we limit the analysis to only the relationships between the focal disciplines and their 10% most cited counterparts. By doing this, we aim to check if our framework will be able to unveil different patterns of IKF in distinct citation classes. The $R^2$ value is reduced in all three plots (Figure 4. d-f), especially for intensity and broadness with $R^2$ close to zero. The broadness and intensity indicator become orthogonal to each other in the highest citation group. This shows that, for the most prominent interdisciplinary relationships, discipline pairs with high broadness are not necessarily characterized by high intensity, which shows the value of considering both broadness and intensity to analyze IKF.

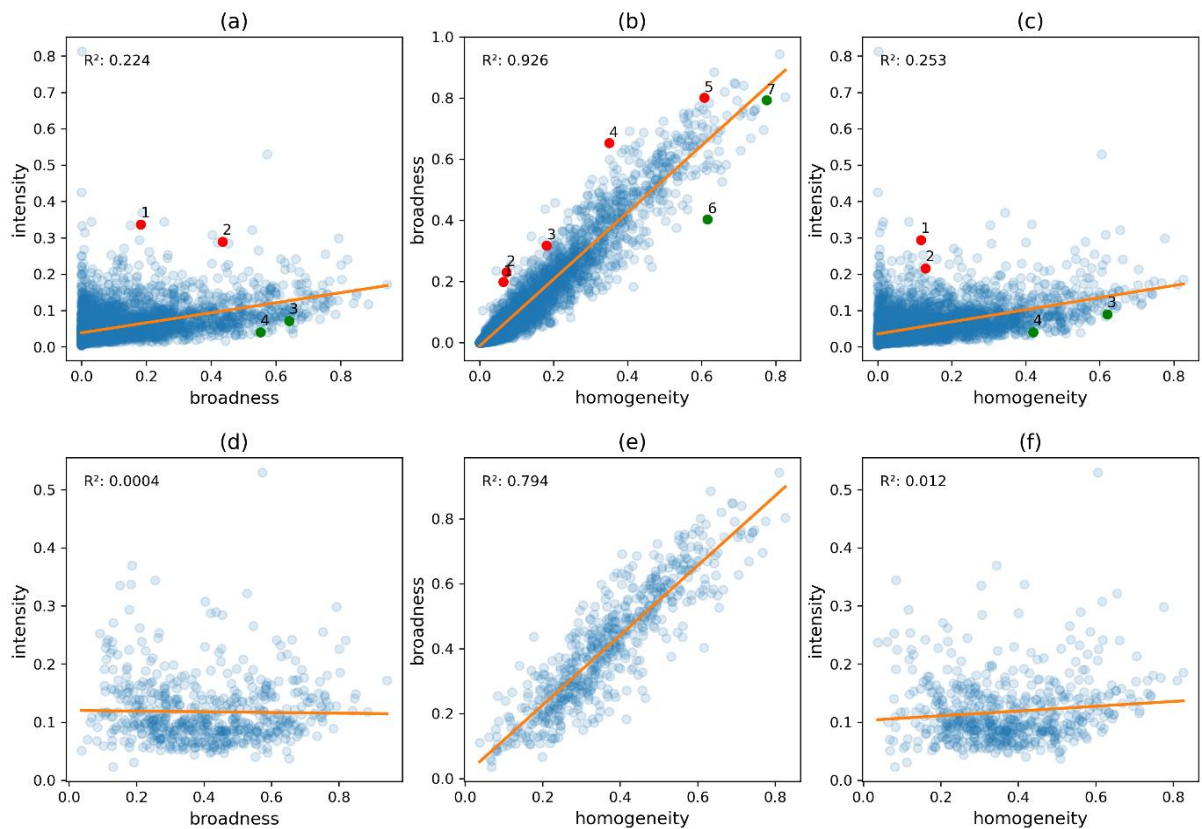

Figure 4. Relationship among the three aspects of IKF. (a-c). all discipline pairs; (d-f). discipline pairs with 10% highest citation percentile. The numbered data points in the upper panel are discussed in Table 1. Linear fits for each subplot and their $R^2$ are visualized and reported.

Table 1. Case studies for selected discipline pairs from Figure 4

### Figure (a)

- Lower broadness & Higher intensity

| No. | Citing discipline | Cited discipline |
|---|---|---|
| 1 | Arts & Design | Psychology & Behavioral Sciences |
| 2 | Law | Psychology & Behavioral Sciences |

- Higher broadness & Lower intensity

| No. | Citing discipline | Cited discipline |
|---|---|---|
| 3 | Physical Chemistry | Applied Physics |
| 4 | Polymer Science | Applied Physics |

**Figure (b)**

- Lower homogeneity & Higher broadness

| No. | Citing discipline | Cited discipline |
|---|---|---|
| 1 | Genetics & Developmental Biology | Applied Mathematics |
| 2 | Pure & Applied Ecology | Applied Mathematics |
| 3 | Genetics & Developmental Biology | Computer Science/Information Technology |
| 4 | Pure & Applied Ecology | Multidisciplinary Biology |
| 5 | Particle & Nuclear Physics | Multidisciplinary Physics |

- Higher homogeneity & Lower broadness

| No. | Citing discipline | Cited discipline |
|---|---|---|
| 6 | Architecture | General & Traditional Engineering |
| 7 | History & Archaeology | Geosciences & Technology |

**Figure (c)**

- Lower homogeneity & Higher intensity

| No. | Citing discipline | Cited discipline |
|---|---|---|
| 1 | Biomaterials & Bioengineering | Dentistry |
| 2 | Literature | History & Archaeology |

- Higher homogeneity & Lower intensity

| No. | Citing discipline | Cited discipline |
|---|---|---|
| 3 | Polymer Science | Physical Chemistry |
| 4 | Polymer Science | Applied Physics |

6.1.2 Relationship between the three aspects of IKF and the volume of citation flow

We further examine the relationships between the proposed three indicators and the volume of citation flow (counts) for discipline pairs. The results are shown in Figure 5a, where the corresponding value for each discipline pair is displayed in scatter plots. An OLS linear fit is conducted using the 'regplot' function from the seaborn Python package (Waskom, 2021), and its $R^2$ value is reported in the plots.

Broadness exhibits a monotonous positive relationship with citation volume between discipline pairs, with 55.4% variance explained. Intensity, however, contributes a significant amount of variations (responsible for only 16% variance) which shows that it has a less correlated relationship with citation volume. The upper outliers with high intensity possibly denote scenarios where the total volume of citations between discipline pairs is small, yet it constitutes an intense citing environment. In other words, it signals a narrow yet intense relationship between disciplines in which a small branch of publications from the citing discipline engages intensively with the knowledge from the cited discipline. When publications from this branch cite them, publications from the cited discipline account for a significant proportion in the reference which is not proportional to expectation (average percentage). Homogeneity shows a monotonous positive relationship with citation volume, which makes intuitive sense since disciplines are more likely to import knowledge from counterparts with similar knowledge bases.

### a. All citation

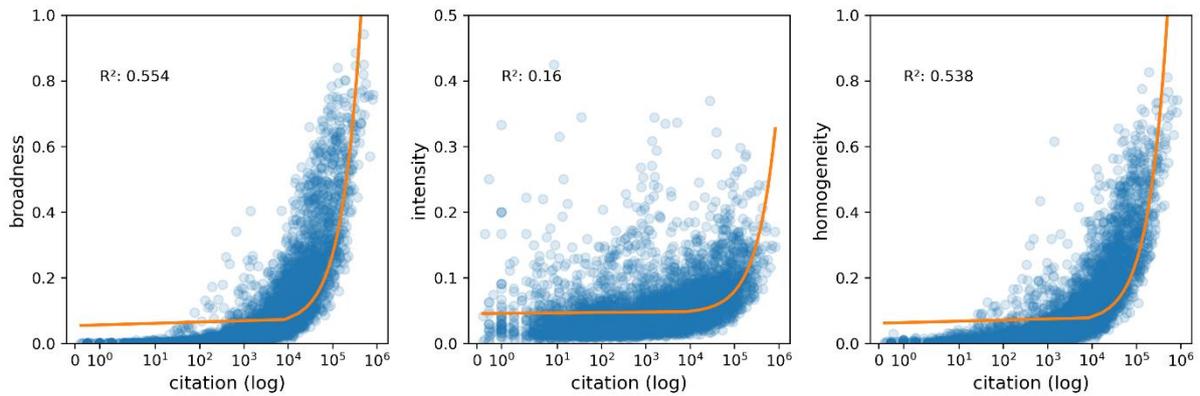

### b. Top 10% citation

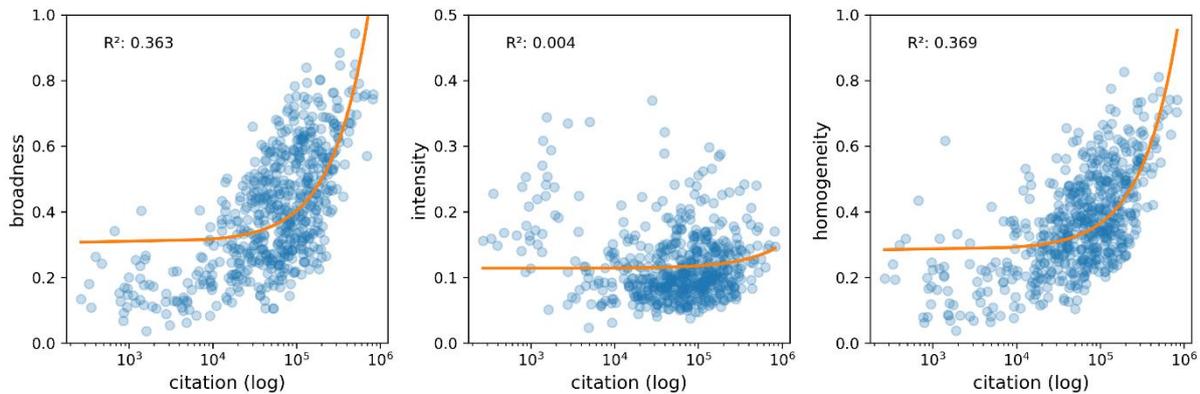

**Figure 5. Relationship between the three indicators and the volume of citation flow. Linear fits for each subplot and their $R^2$ are visualized and reported. (a) relationship for all discipline pairs; (b) relationships for discipline pairs with top 10% citation volumes.**

We once again examine discipline pairwise relationships in the top 10% citation group separately and arrive at similar findings as to the previous one. In the 10% citation group (Figure 5b), all three aspects are found to be less correlated with citation volume, as indicated by lower values in $R^2$. This shows that our proposed indicators possess greater discriminative power among discipline relationships with the largest volume of citations, which are normally the foci in the study of interdisciplinarity.

*6.2 Use cases*

To demonstrate how to use the recognized three aspects of IKF for quantitative science studies or research evaluation, we test the capability of our proposed framework through two use cases. The first use case utilizes this framework to investigate and demonstrate unique patterns of IKF for disciplines. The second use case aims to indicate the relative interdisciplinarity of scientific journals from Library and Information Science (LIS) in comparison to the LIS field at large and visualize the process of knowledge integration.

6.2.1 Indicating the interdisciplinarity of disciplines

Here we choose eight disciplines as examples to illustrate how the framework can be employed to unveil different patterns of IKF and interdisciplinarity. They are (1) Applied Mathematics, (2) Business, Economics, Planning, (3) Cell Biology, (4) Computer Science/Information Technology, (5) Environmental Science & Technology, (6) General & Traditional Engineering, (7) Immunology, and (8) Linguistics. The scatterplots (Figure 6) illustrate how the focal discipline (indicated in the subtitle) cites the other disciplines under our framework. Below we provide two discussions on how the proposal characterizes the interdisciplinarity of disciplines.

**Indicating the degree or status of interdisciplinarity.** In general, all disciplines rely heavily on a few disciplines in citing behavior, with more extreme values for all three aspects of IKF. Nevertheless, disciplinary discrepancies can be spotted. Some disciplines, for instance, Computer Science/Information Technology, Linguistics, and Cell Biology have a significantly stronger relationship with only one discipline for all three aspects. On the other hand, disciplines like Immunology and General & Traditional Engineering have a more balanced relationship with several disciplines, shown by a more scattered distribution. In addition, some disciplines, e.g., Business, Economics, Planning and Cell Biology, exhibit high homogeneity in the knowledge base with many disciplines, despite low broadness or intensity with them. One possible interpretation is that these disciplines study topics that shares significant cognitive components with many disciplines, yet prefer to conduct research in a more insular or independent way. They may also have nurtured various interdisciplinary topics within themselves, hence less necessity in citing other disciplines directly. Potentially, if one desires an aggregated indicator for interdisciplinarity under our framework, the degree of variation (or scattering) of data points in these plots could be a candidate solution. For instance, we calculate the sum of standard deviations of three indicators for each citing discipline and correlate it with the median interdisciplinarity values for all publications in it, namely true diversity or TD (Zhang et al., 2016) and DIV (Leydesdorff et al., 2019). The Pearson coefficients between the sum of standard deviations and some interdisciplinarity indicators are relatively high: 0.777 for DIV and 0.680 for TD, which indicates a relatively strong positive relationship between them (see SI A for scatter plot). Our goal in this analysis is not to propose a new indicator of interdisciplinarity, but to see the connection between our proposed framework and established indicators.

**Characterizing different types of disciplinary relationships.** Another valuable insight that can be drawn from here is the distinct patterns of IKF in pairwise discipline relationships. For Business, Economics, Planning, its relationships with Psychology & Behavioral Sciences, and Applied Mathematics are apparently different in broadness and intensity. Psychology & Behavioral Sciences appear to possess a more intense impact on the focal discipline, while Applied Mathematics more broadly influences its research. Applied Mathematics has become an indispensable and pervasive component in economics research, but it does not occupy a large proportion of references. In contrast, Psychology & Behavioral Sciences only relates closely to some dedicated branches of Business, Economics, Planning (e.g. marketing) yet contributes a significant amount of knowledge. Such patterns can also be found in Applied Mathematics, Environmental Science & Technology, and Linguistics, whose relationships with two disciplines (highlighted dots) have opposite trade-offs on broadness and intensity. This corroborates with the orthogonality between broadness and intensity in the higher 10% citation group. In addition, with a similar cognitive overlap in the knowledge base (similar homogeneity), some disciplines are broadly/intensively cited while some are not. For instance, Applied Mathematics shares most references with Computer Science/Information Technology (H = 0.51) and General & Traditional Engineering (H=0.45). Yet the former has a broadness of 0.57, whereas that of the latter is only 0.32.

Using our framework, one can uncover more contextual and unique features of IKF and interdisciplinarity than a single indicator like citation counts or PCTC from previous studies. We would also like to point out what our framework can contribute to established interdisciplinarity indicators. For instance, Cell Biology and Immunology are associated with similar high interdisciplinarity values using existing indicators. However, the former's high interdisciplinarity is associated with strong extramural citations with a limited number of disciplines, while the latter cites more widely and impartially from a cluster of disciplines. By using both established indicators and our indicating process, one can both quickly assess the interdisciplinarity of entities and infer the reason in detail.

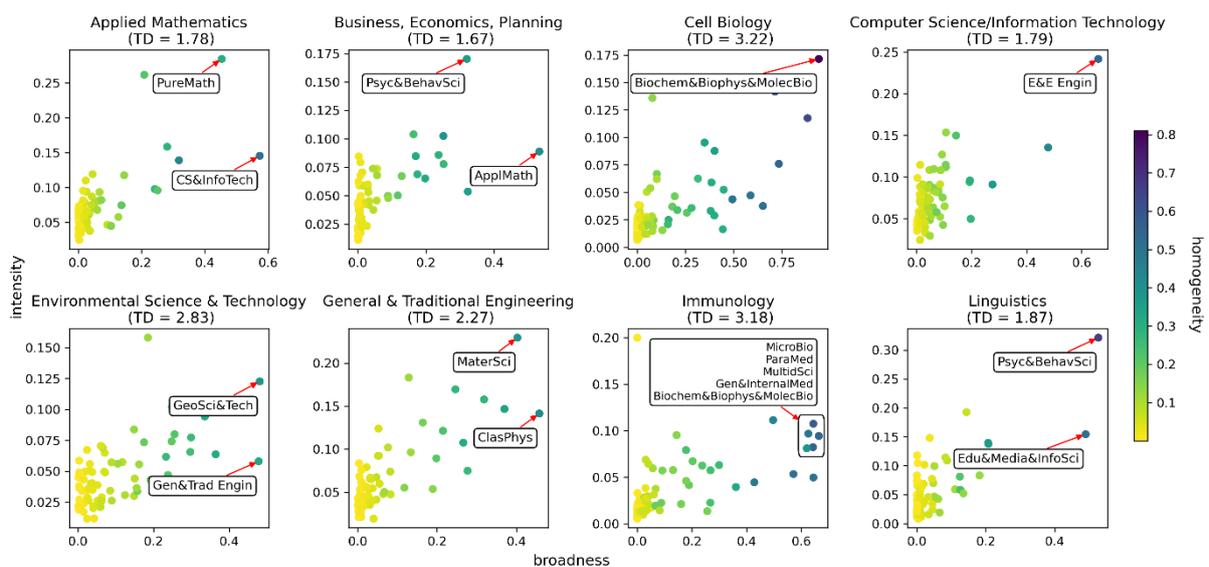

**Figure 6. IKF of disciplines under our framework.** These scatterplots illustrate how the focal discipline (indicated in the subtitle) cites the other disciplines (denoted as dots) under our framework. Citations toward the discipline itself

are not included. Some special data points are highlighted and identified. The interdisciplinarity value (TD, True diversity) is quantified and reported for each discipline as the median TD value for all its publication. See a 3-D projection version in SI.

6.2.2 Indicating the relative interdisciplinarity of journals

To further demonstrate the utility of our framework, we examined the relative IKF of selected scientific journals in LIS under our framework in comparison to LIS publications in general, in other words, how certain LIS journals cite other disciplines differently from LIS at large. We choose 87 journals from JCR 2019 categorized under the field "Information Science & Library Science" as representative of the LIS field and select seven journals as examples for study, namely *Information and Organization* (IO), *Information Processing & Management* (IPM), *Journal of the Association for Information Science and Technology* (JASIST), *Journal of Informetrics* (JOI), *MIS Quarterly* (MISQ), *Research Evaluation* (RE), and *Scientometrics* (SCIM).

For the calculation of three indicators $D(J, Y)$, we set the citing side as one of the journals ($J_i$) or all journals from LIS in our dataset ($J_{all}$) and cited side as cited disciplines ($Y$). We then calculate the difference for each indicator between the focal journal and the LIS field as $D_{diff}(J_i, Y)$, where:

$$D_{diff}(J_i, Y) = D(J_i, Y) - D(J_{all}, Y) \qquad (4)$$

In doing so, we aim to elucidate how the knowledge base of these journals deviates from the norm of the field of origin and try to describe the distinctive nuance in their knowledge portfolios.

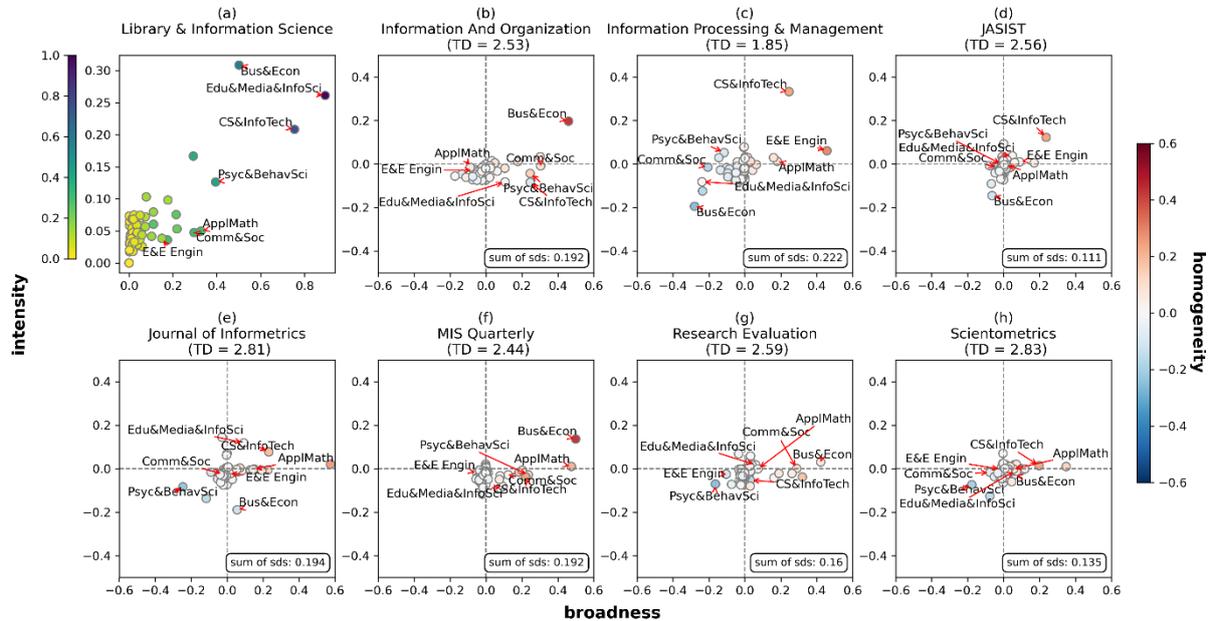

Figure 7. A case study of LIS journals. (a). The IKF of the LIS field at large under our framework; (b-h). The relative IKF of seven LIS journals under our framework, compared to the LIS field. The sum of standard deviations of three indicators for each journal is reported in the bottom-right box as an indication of deviation. The interdisciplinarity value (TD) is quantified and reported for each journal as the median TD value for all its publication.

Figure 7a illustrates the IKF of the LIS field under our framework; Figure 7 b-h showcases the deviations of the knowledge base (relative IKF) for seven LIS journals from their field of origin. A positive value in each of the aspects of IKF in b-h means the focal journal has a stronger relationship or greater cognitive affinity with the cited discipline than the LIS field in general. The more extreme the value, the greater the focal journal deviates from LIS as a whole regarding interdisciplinary engagements. Seven cited disciplines are annotated as examples, namely:

- Applied Mathematics (ApplMath)
- Business, Economics, Planning (Bus&Econ)
- Computer Science/Information Technology (CS&InfoTech)
- Community & Social Issues (Comm&Soc)
- Electrical & Electronic Engineering (E&E Engin)
- Education, Media & Information Science (Edu&Media&InfoSci)
- Psychology & Behavioral Sciences (Psyc&BehavSci)

We now discuss the differences between the seven journals regarding their relationships (deviations) with these seven cited disciplines.

IPM, with a technological orientation, is not surprisingly found to possess stronger relationships with Computer Science/Information Technology in all three aspects of IKF, and so do JASIST, JOI, and SCIM. The broadness of the aforementioned four journals is close to 100%, which means almost all papers from these journals cite Computer Science/Information Technology. The intensity, on the other hand, varies among them with IPM possessing the highest intensity value while JASIST, JOI, and SCIM exhibit intensity values in descending order. IPM is more cognitively homogenous with Electrical & Electronic Engineering and cites it with a greater broadness and intensity, which is lacking in the rest of the journals. Its relationship (especially broadness and homogeneity) with the four social science disciplines is weaker compared to other journals.

In contrast, MISQ clearly has a stronger connection with social science disciplines in broadness. Particularly, MISQ cites Business, Economics, Planning more heavily in all three aspects (same for IO). But it is the only journal in our selected examples that cite Business, Economics, Planning heavily in conjunction with Applied Mathematics in great broadness, characterizing one of the most unique features of MISQ. A greater percentage of MISQ publications cite Applied Mathematics and exhibit greater homogeneity in the knowledge base than the general LIS fields but in a similar intensity.

IO, as a journal with a focus on the relationship between information technology and social organizations, also shares a greater affinity to several social science disciplines and less affinity to STEM disciplines such as Applied mathematics and Electrical & Electronic Engineering. IO is the only journal in the selected seven that cites Applied Mathematics in lower broadness than average.

Another journal that possesses higher IKF values with Business, Economics, Planning for all three indicators is RE, although smaller than that of MISQ and IO. It has a stronger connection with other social science disciplines, for instance, Community & Social Issues,

Philosophy & Religion, and Political Science & Administration, instead of the other two we have labeled in Figure 7b. RE exhibits less homogeneity in the knowledge base with Computer Science & Information Technology, similar to IO.

JASIST and SCIM are the closest to the LIS field in terms of knowledge base and IKF, which is also shown by the lowest value for the sum of standard deviations of the three aspects of IKF. Their relationships with most of the disciplines are not quite different from that of the LIS field. They both cite Computer Science/Information Technology in a greater broadness but with a greater intensity only for JASIST. SCIM cites Psychology & Behavioral Sciences with a smaller intensity, whereas JASIST cites Business, Economics, Planning with a smaller intensity. JASIST's knowledge base is most similar to the knowledge base of the general LIS field, which makes intuitive sense since it is designed to cover a wide range of topics in LIS.

Finally, JOI shares a similar knowledge portfolio with JASIST and SCIM, but with a few discrepancies. It cites Multidisciplinary Sciences (the right-most point) at a significant level of broadness and homogeneity, which is absent in other journals. Additionally, Education, Media & Information Science appears to be a vital knowledge source with greater intensity for JOI, yet that of both Psychology & Behavioral Sciences and Business, Economics, Planning are lower than the general LIS field.

Through this case study on the relative IKF of journals, we demonstrate how our proposed methods of characterizing IKF can vividly indicate the interdisciplinarity of the knowledge base for a certain entity, which is lacking in aggregated indicators. For instance, existing indicators assign similar values of interdisciplinarity to journals such as IO and JASIST, although they exhibit quite different characteristics in relative interdisciplinary contexts. We once again calculate the Pearson correlation between interdisciplinarity values and the sum of standard deviations for the difference of three indicators compared to the LIS at large (in Figure 5 b-h) and arrived at a coefficient of -0.093 (see SI. A for details). This shows that the relative IKF of LIS journals, the deviations in interdisciplinarity from LIS (relative interdisciplinarity), is different from its interdisciplinarity in a global context; that is to say, a journal that is found to be more interdisciplinary does not necessarily deviate significantly from its disciplinary norm. This illustrates the necessity of our proposed approach in relative terms that can unveil the relative interdisciplinarity of entities in comparison with their peers. Such operationalization of relative IKF can highlight some of the most significant interdisciplinary features of entities and facilitate valid peer comparisons.

## 7. DISCUSSION AND CONCLUSION

This paper proposes a new perspective quantifying the unexplored aspects of interdisciplinary knowledge flow to infer the pattern and dynamics of interdisciplinarity. These three aspects of IKF are broadness, intensity, and homogeneity, operationalized using three indicators. Using this conceptualization, we manage to offer a contextual and holistic understanding of the knowledge exchanges between disciplines and provide more details and scrutiny than simple citation counts. In order to validate this framework, we investigate the relationships among the three aspects for discipline pairs as well as the relationships between the three aspects and citation counts. We showcase that our proposed method can capture distinct aspects of IKF, especially in the highest citation groups. As two use cases, we apply this method to examine the disparate patterns of IKF in academic disciplines and scientific journals. We argue that our method carries great power and potential in indicating interdisciplinarity and avoids some of the problems of composite indicators of interdisciplinarity.

The inspiration for this article is triggered by our previous research (Zhou et al., 2021, 2022), in which we studied the evolution of interdisciplinarity in the social sciences using several aggregated interdisciplinarity indicators. The study offers a succinct and general perspective on interdisciplinarity by approaching it as diversity in the knowledge base, which can be subsumed into a single measurement. In the task of finding a proxy to approximate interdisciplinarity, it is essential that one pursues simplicity and accuracy, by which we mean to improve the indicators of interdisciplinarity until it is closest to the ideal definition of interdisciplinarity. On the other hand, we should also realize that it is perhaps a "mission impossible" in the first place; complexity and comprehensiveness also matter in certain scenarios. Rafols (2020a) stresses the importance of "directionality", i.e. the orientation of research contents in science and innovation. In the context of interdisciplinarity, we argue that an equally important question, if not more, could be "what is inter-disciplined" compared to our current objective to quantify "how interdisciplinary". Most attention and efforts have been directed to the latter question. This article attempts to offer a tentative solution to the less explored question, seeking to propose and utilize several indicators to arrive at disaggregated and more contextual understandings of interdisciplinarity. We would also argue that it's unnecessary to completely abandon interdisciplinarity indicators, although they are currently under criticism. A combination of both indicators and the indicating processes of interdisciplinarity can deliver better and more accurate implications, just like what we did in this study to report both interdisciplinarity values and the three aspects of IKF at the same time.

Furthermore, besides its kaleidoscopic connotations on the conceptual level, we argue that interdisciplinarity is a multifaceted construct also in the context of research evaluation and management that exists in different forms, delivers different interests to different parties, and should be approached by different methodologies. For instance, funding institutions adopt various policy tools to facilitate interdisciplinary research in different contexts and should be provided with suitable empirical evidence to assist decision-making. To begin with, evaluations of interdisciplinarity can be embedded in performance-based research funding allocation schemes as one of the factors that will affect the amount of funding each university obtain. In Flanders, the Dutch-speaking region of Belgium, a parameter for interdisciplinary research is scheduled to be added to the funding partition formula from 2024 onwards (Luwel, 2021). Under such a scenario, quantitative indicators could be more suitable to ensure a tangible and relatively unbiased evaluation process and demonstrate the funding body's encouragement of interdisciplinary research. On the other hand, the review process for interdisciplinary research projects commonly involves a panel of referee experts (evaluation committee) and requires an abductive and participatory approach to evaluate interdisciplinarity. At Ghent University in Flanders, for example, this implies that applications for interdisciplinary research projects are firstly pre-selected and evaluated on the "level of interdisciplinarity" of the proposals[1]. Three criteria for interdisciplinarity are stipulated: dissimilarity in involved disciplines or expertise, equally essential and integrated inputs from involved disciplines, and potential for the development of a new field of study or new insights in both disciplines. Under such a scenario, we suggest that interdisciplinarity is better evaluated and presented in a contextual process where the decision-making process is supported by quantitative and qualitative background material: e.g. interdisciplinary profiles at the international level; disciplinary and interdisciplinary strengths, weaknesses, and planning at the national level; existing disciplinary and interdisciplinary strengths and outputs at the university level. The reviewing process can therefore be more informative and be situated in grander initiatives such as international competition, national R&D strategic planning, and university management.

On the other hand, as more funding mechanisms are requiring interdisciplinary elements in the research design of applications, funding bodies should also be vigilant to prevent inflationary claims of interdisciplinarity wherever the proposed research is already subsumed under disciplinary practices. Initial filtering on applications aided by information tools can perhaps partly offset such inflationary tendencies and reduce the cost of evaluation. Besides research evaluation, funding bodies can also play important roles in "identifying questions that need an interdisciplinary approach",  "launching and shaping initiatives" (funding), and "establishing the architecture of an interdisciplinary programme" or research centers (Marsden et al., 2011).

By proposing our method capturing various aspects of interdisciplinary knowledge flow, we aim to contribute another scientometric tool to research evaluation practices that require a contextual understanding of the knowledge portfolio of scientific entities, not only disciplines and journals, but also research teams, research institutes, countries, or other groups of entities. We would also like to call for more diverse quantitative tools for indicating interdisciplinarity to cope with the diverse policy needs in facilitating interdisciplinary research in research evaluation and management.

Our study has several limitations and room for future studies. First, we focus on the citing side of disciplines or journals, i.e. references, which only entails one side of the interdisciplinary knowledge flow. Since the framework is designed to capture asymmetric relationships between entities, in future studies, we will also investigate the cited side of knowledge flow and unveil the dynamics of knowledge diffusion under our framework. In addition, in order to propose and validate this model first, we conduct analyses from a static perspective that studies the relationships between disciplines within a set time frame. Temporal evolutions of interdisciplinarity under our model will be explored in a future study.

---

[1] https://www.ugent.be/en/research/funding/bof/iop

# ACKNOWLEDGEMENTS

We thank the Flemish Government for its support to the Centre for R&D Monitoring (ECOOM), which made possible this research as part of the program on the study of interdisciplinarity and impact. The opinions in the paper are the authors' and not necessarily those of the government.

We thank Ying Huang, Ronald Rousseau, and Lin Zhang for their valuable comments. We appreciate fruitful and inspiring discussions with our colleagues at ECOOM-UAntwerpen. We thank Ludo Waltman and the other two reviewers for their constructive comments which significantly improved this paper.

# SUPPLEMENTARY INFORMATION (SI)

## SI. A. Additional plots mentioned in the article.

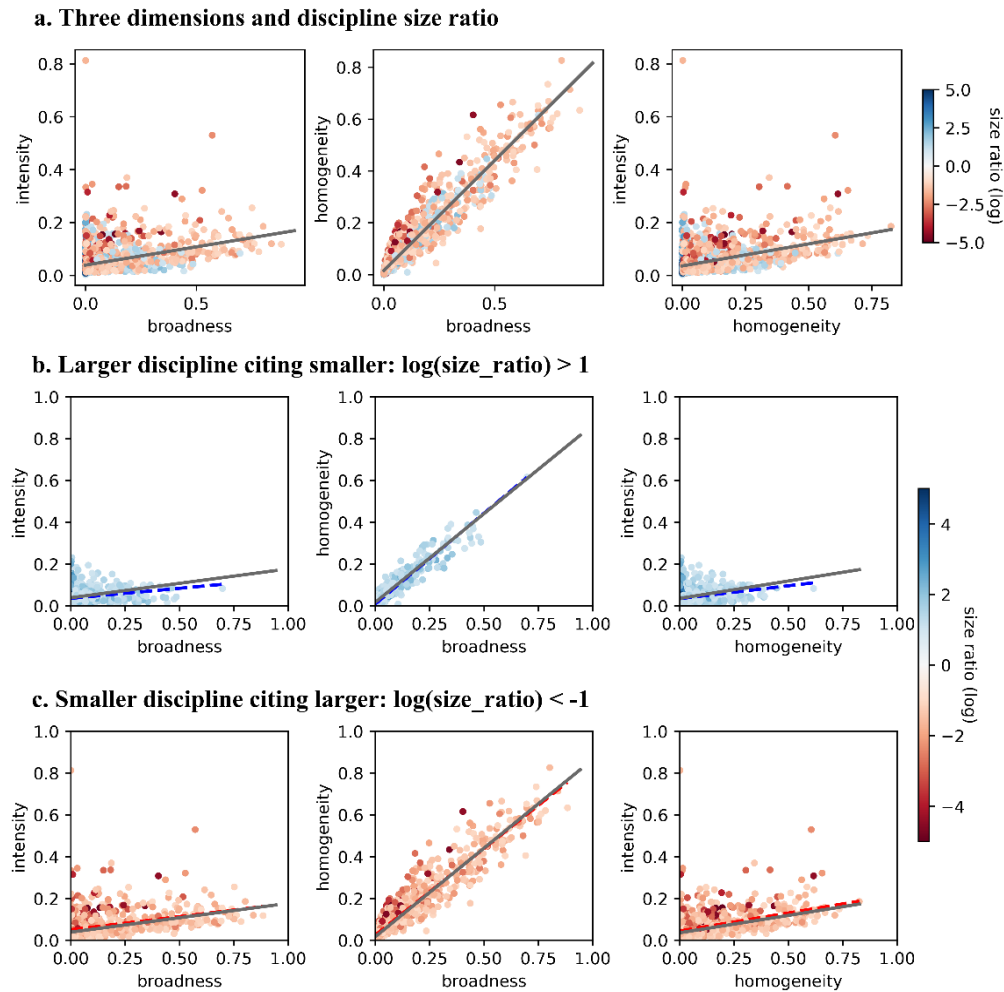

**Figure A1. Considering discipline size in the relationship between aspects of IKF.** The size ratio is quantified as the logarithmic ratio of the number of publications for the citing discipline and the cited discipline, denoted by colors – blue if the citing discipline is larger ($\log size\_ratio > 0$), red if cited discipline is larger ($\log size\_ratio < 0$). (a) relationships between three indicators and discipline size ratio for all discipline pairs. (b-c) Disciplines pairs with significant size imbalances are selected ($|\log size\_ratio| > 1$). Linear fits for all discipline pairs (grey solid line) and selected ones (colored dashed line) are reported.

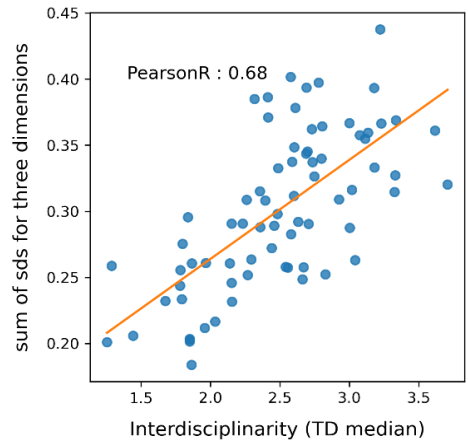

**Figure A2. Relationship between the "scattering" of three indicators and the interdisciplinarity value (TD median)**

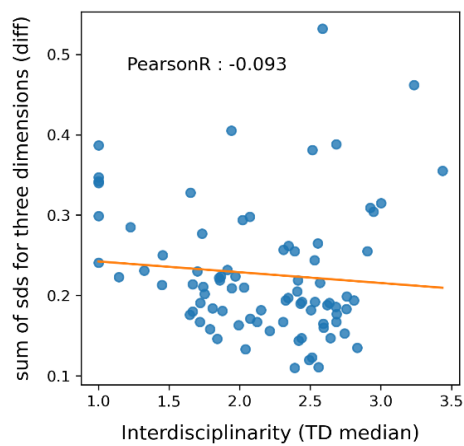

**Figure A3. Relationship between the "scattering" of three indicators in relative format and interdisciplinarity value (TD median)**

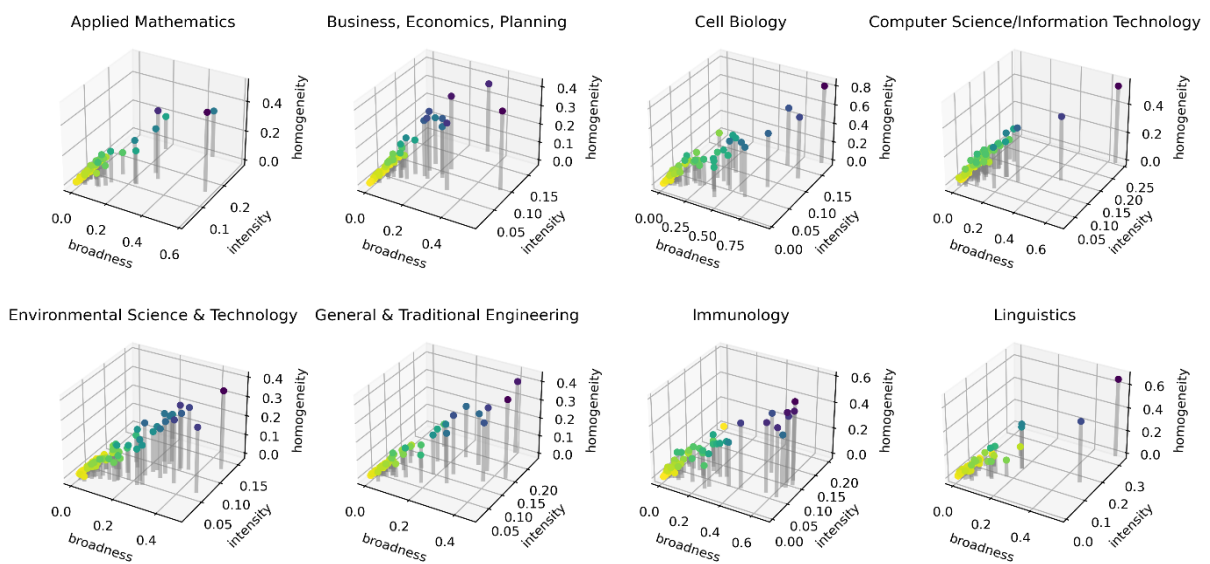

**Figure A4. 3-D projection view on the relationships among the three aspects of IKF**

# SI. B. PROPERTIES OF THE PROPOSED THREE INDICATORS OF IKF

**B1. Monotonicity**

B1.1 Broadness

The broadness of IKF of $X$ citing $Y$ is defined as a fraction as follows,

$$B(X,Y) = \frac{\sum_{i \in X} \delta_i}{|X|} = \frac{|X'|}{|X|}$$

where $X'$ denotes a randomly selected subset of $X$ that cites $Y$. Having the size of $X$ fixed, $B(X,Y)$ increases monotonically with $X'$. On the other hand, having the size of $X'$ fixed, $B(X,Y)$ decreases monotonically with $|X|$.

B1.2 Intensity

The intensity of IKF of $X$ citing $Y$ is defined as a fraction as follows,

$$I(X,Y) = \frac{\sum_{i \in X, j \in Y} M_{ij}}{\sum_{i \in X, j=1,\ldots,n}(M_{ij}\delta_i)}$$

$$= \frac{Cit(X,Y)}{\sum Cit(X',d)} = \frac{Cit(X',Y)}{\sum Cit(X',d)}$$

where a function $Cit$ quantifies the number of citations from one publication set to another one. The numerator $Cit(X,Y)$ is the number of citations from $X$ to $Y$, equal to $Cit(X',Y)$, whereas the denominator is the total number of outward citations (to all other entities $d$) by $X$. Having $\sum Cit(X',d)$ fixed, $I(X,Y)$ increases monotonically with $Cit(X',Y)$. On the other hand, having the numerator fixed, $I(X,Y)$ decreases monotonically with the denominator.

B1.3 Homogeneity

The homogeneity of $X$ with $Y$ is defined as a fraction as follows:

$$H(X,Y) = \frac{\sum_{i \in X, \gamma=1,\ldots,n} M_{i\gamma}\varphi_{\gamma,Y}}{\sum_{i \in X, j=1,\ldots,n} M_{ij}}$$

Having the number of references of $X$ fixed (denominator), an increasing number of co-cited references ($M_{i\gamma}\varphi_{\gamma,Y}$) between $X$ and $Y$ in the numerator is associated with higher $H(X,Y)$.

**B2. Size independence**

B2.1 Definition

Here, size independence means that, for entity A, an identical subset of A with only a difference in size should obtain the same values for the three indicators.

B2.2 Mathematical proof

**Broadness**

Following the definition in 2.1, for entity X, we define an identical subset of $X$ named $X_s$. According to 2.1 and the definition of broadness, we should have:

$$|X_s| = \alpha|X| \quad and \quad |X'_s| = \alpha|X'|, \quad (0 < \alpha \leq 1)$$

Then we have:

$$B(X, Y) = \frac{|X'|}{|X|}$$

and,

$$B(X_s, Y) = \frac{\alpha|X'|}{\alpha|X|} = \frac{|X'|}{|X|}$$

hence,

$$B(X, Y) = (X_s, Y)$$

We conclude that $B(X, Y)$ is independent of $|X|$.

**Intensity**

Similar to 2.2.1, we conclude that $I(X, Y)$ is independent of $|X|$.

**Homogeneity**

We have:

$$H(X, Y) = \frac{\sum_{i \in X, \gamma = 1, \ldots, n} M_{i\gamma} \varphi_{\gamma, Y}}{\sum_{i \in X, j = 1, \ldots, n} M_{ij}}$$

We divide $n$ for both the numerator and the denominator of the above equation and obtain:

$$H(X, Y) = \frac{\sum_{i \in X, \gamma = 1, \ldots, n} \frac{M_{i\gamma} \varphi_{\gamma, Y}}{n}}{\sum_{i \in X, j = 1, \ldots, n} \frac{M_{ij}}{n}} = \frac{\sum_{i \in X, \gamma = 1, \ldots, n} p_{XY}}{\sum_{i \in X, j = 1, \ldots, n} p_{X \rightarrow}}$$

Where $p_{X \rightarrow}$ and $p_{XY}$ are the probabilities that publications in X cite other publications and that publications in X cite publications that are also cited by publications in Y.

On the other hand, we have:

$$H(X_s, Y) = \frac{\sum_{i \in X_s, \gamma = 1,\ldots,n} \frac{M_{i\gamma} \varphi_{\gamma,Y}}{n}}{\frac{\sum_{i \in X_s, j = 1,\ldots,n} M_{ij}}{n}} = \frac{\sum_{i \in X_s, \gamma = 1,\ldots,n} p_{X_s Y}}{\sum_{i \in X_s, j = 1,\ldots,n} p_{X_s \to}}$$

Where, again, $p_{X_s Y}$ and $p_{X_s \to}$ correspond to the probabilities that publications in $X_s$ cite publications that are also cited by Y and that publications $X_s$ cite other publications.

As $X_s$ is a randomly selected sub-set from $X$, we have $p_{X_s Y} = p_{XY}$ and $p_{X_s \to} = p_{X \to}$. We then have $H(X, Y) = H(X_s, Y)$. That is, homogeneity is also size independent.

B2.3 Empirical tests

We further test the property of size independence in empirical settings. For each of the 74 disciplines in our dataset, we randomly select 10%, 20%, …, 80%, and 90% of their publications to represent a subset of that discipline. We then calculate the three indicators for each discipline pair in each random subset and quantify the difference between simulated results and with original results using the full set. We plot the average differences for each indicator and each random trial. The random process is repeated 10 times and the distribution of average differences is shown in Figure B1.

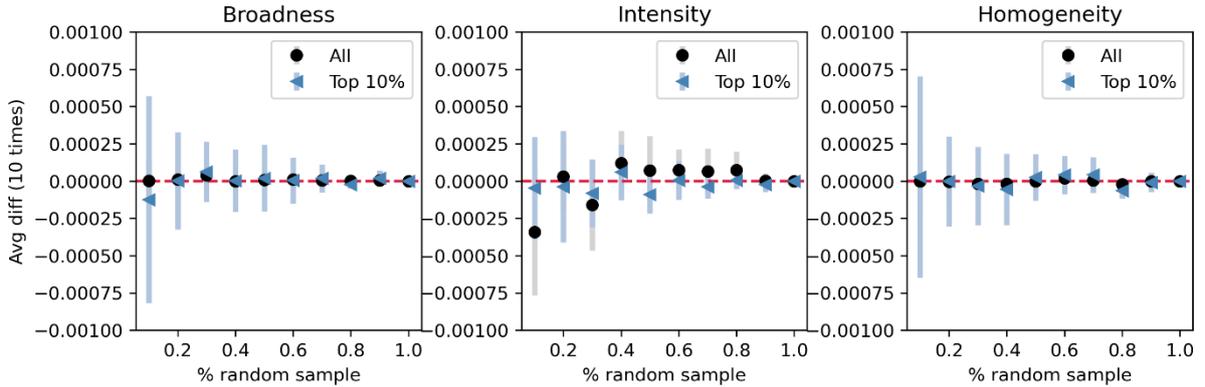

Figure B1. The average difference in the value of three indicators between the full set and randomly simulated subsets. The back circle denotes the results for all discipline pairs, while the blue triangle denotes that of 10% of discipline pairs with the greatest citations between them. The x-axis denotes different sample sizes (percentage of the full set), while the y-axis denotes the difference between results from the full set and random samples. The mean difference of ten random trials is shown as the black dots and the grey bar represents the standard deviation. A horizontal red dashed line is shown to represent y=0, i.e., no difference between the full set and random subsets, hence size independence.

We find that all three indicators are independent of the size of citing entities since values for various subsets of the entity remain around zero with minor fluctuations. So if a discipline has $n$ publications with a broadness (same for intensity and homogeneity) distribution of $D$, a random sample of that discipline having $0.5n$ publications is also expected to follow $D$.

# SI. C. RELATIONS BETWEEN THE PROPOSED METHOD AND CURRENT SCIENTOMETRIC INDICATORS FOR INTERDISCIPLINARITY

The framework shares some ingredients and conceptualization with the established scientometrics indicators, including both knowledge-flow-based (Porter & Chubin, 1985) and diversity-based indicators (Rafols & Meyer, 2010). The calculation for both types of indicators is centered around the proportion/volume of citations from the entity of interest to other fields, which is equivalent to PCTC and $p_i$ in diversity-based IDR indicators (e.g., Rao-Stirling diversity in equation C1). In our framework, we further decompose $p_i$ to two aspects, namely broadness and intensity, to provide a detailed lens for capturing interdisciplinary knowledge flow. The second ingredient that some diversity-based IDR indicators employed is the dissimilarity between disciplines, i.e. disparity ($d_{ij}$ in equation 5). We also include this aspect since it offers insights on the cognitive proximity between disciplines. Previous studies have employed several measurements for disparity such as bibliographic coupling, co-citation, and cross-citation. We try to indicate disparity/similarity using the proposed homogeneity indicator, which is also a form of bibliographic coupling. The differences between ours and existing ones are twofold: we explicitly designed our framework as asymmetrical while previous disparity indicators yield the same results for A citing B and B citing A; the second difference is that we purposely design this indicator in a simple, intuitive and empirically meaningful form, which is the degree of overlap in the knowledge base.

$$RS_{IDR} = \sum_{i,j(i \neq j)} d_{ij}(p_i p_j) \tag{C1}$$

Our framework contains all the elements in previous indicators on IDR. The framework directly modifies the knowledge-flow-based indicators on interdisciplinarity. For diversity-based interdisciplinarity, three elements of interdisciplinarity, namely variety, balance, and disparity, are also explicitly or implicitly indicated in our framework with improvements. 1) Variety is equivalent to the number of non-zero elements in the array of broadness and intensity, which is the number of disciplines cited by the evaluated entity. Our approach offers more information on the variety in the sense that the volume of citations is also depicted. In addition, contrary to variety, which increases with even only one citation to a new extramural field, our approach is less sensitive to minor changes. 2) Balance is implicitly indicated by the distribution of broadness and intensity: a skewed distribution over fields corresponds to a lower balance and a uniform distribution shows a higher balance. 3) Disparity, or dissimilarity between disciplines, is explicitly included in our framework as homogeneity/heterogeneity which quantifies the overlap/difference in the knowledge base of the entity of interest and another field.